\documentclass{acmart} 
\usepackage{enumerate}

\AtBeginDocument{%
  \providecommand\BibTeX{{%
    \normalfont B\kern-0.5em{\scshape i\kern-0.25em b}\kern-0.8em\TeX}}}

\copyrightyear{2022}
\acmYear{2022}
\setcopyright{rightsretained}
\acmConference[FAccT '22]{2022 ACM Conference on Fairness, Accountability, and Transparency}{June 21--24, 2022}{Seoul, Republic of Korea}
\acmBooktitle{2022 ACM Conference on Fairness, Accountability, and Transparency (FAccT '22), June 21--24, 2022, Seoul, Republic of Korea}\acmDOI{10.1145/3531146.3533143}
\acmISBN{978-1-4503-9352-2/22/06}

\usepackage{multirow}

\begin{document}

\title{Behavioral Use Licensing for Responsible AI}

\author{Danish Contractor}
\authornote{Member of the Responsible AI Licenses (RAIL) Initiative}
\affiliation{\institution{IBM Research AI}
\country{India}}
\email{danish.contractor@ibm.com}

\author{Daniel McDuff}
\authornotemark[1]
\affiliation{\institution{Microsoft Research}
\country{United States of America}}
\email{damcduff@microsoft.com}

\author{Julia Katherine Haines}
\affiliation{\institution{RAIL}
\country{United States of America}}
\email{juliahaines@me.com}

\author{Jenny Lee}
\affiliation{\institution{RAIL}
\country{United States of America}}
\email{jnlee@post.harvard.edu}

\author{Christopher Hines}
\authornotemark[1]
\affiliation{\institution{K\&L Gates LLP}
\country{United States of America}}
\email{christopher.hines@klgates.com}

\author{Brent Hecht}
\authornotemark[1]
\affiliation{\institution{Northwestern University}
\country{United States of America}}
\email{bhecht@northwestern.edu}

\author{Nicholas Vincent}
\affiliation{\institution{Northwestern University}
\country{United States of America}}
\email{nickvincent@u.northwestern.edu}

\author{Hanlin Li }
\affiliation{\institution{Northwestern University}
\country{United States of America}}
\email{lihanlin@u.northwestern.edu}

\renewcommand{\shortauthors}{Contractor, et al.}



\begin{abstract}
With the growing reliance on artificial intelligence (AI) for many different applications, the sharing of code, data, and models is important to ensure the replicability and democratization of scientific knowledge. 
Many high-profile academic publishing venues expect code and models to be submitted and released with papers.
Furthermore, developers often want to release these assets to encourage development of technology that leverages their frameworks and services. 
A number of organizations have expressed concerns about the inappropriate or irresponsible use of AI and have proposed ethical guidelines around the application of such systems.  While such guidelines can help set norms and shape policy, they are not easily enforceable. 
In this paper, we advocate the use of licensing to enable legally enforceable behavioral use conditions on software and code and provide several case studies that demonstrate the feasibility of behavioral use licensing. We envision how licensing may be implemented in accordance with existing responsible AI guidelines. 
\end{abstract}

\ccsdesc[500]{Social and professional topics~Licensing}

\keywords{AI licensing, ethical guidelines and principles, enforceable mechanisms}

\maketitle




%

\section{Introduction}

Academic communities rely on the sharing of ideas, materials, and data. ``Openness'' of this kind is vital because it allows methods to be verified, saves time and resources, and allows researchers to more effectively build upon ideas that have come before. It lowers the bar to entry for academics and reduces friction in translating research to applications.

Accordingly, openness around code and other intellectual property has become one of the core values of computer science. High profile computer science conferences, such as NeurIPS and CVPR, expect code to be submitted and released with published submissions \cite{NeuralInformationProcessingSystemsFoundation2021Sep} and the policy of some journals is that ``{\em authors are required to make materials, data, code, and associated protocols promptly available to readers without undue qualifications}''~\cite{natureStandards}. Within the human-computer interaction (HCI) research community specifically, there has been discussion of how to support open research practices for HCI \cite{wilson2011replication,transparencyCHI} and how HCI can help other fields practice open computing research \cite{reproHCI}.

Advances in computing have introduced new challenges in maintaining openness. A new generation of end-to-end neural approaches to AI~(e.g., convolutional neural networks, recurrent neural networks, generative adversarial networks, transformers) is leading to systems becoming increasingly generalized, rather than limited to feature sets tailored for specific domains. While AI systems are not limited to neural models, the need for control over use has been accelerated due to advances in model performance. For example, they can generate text~\cite{brown2020language}, images~\cite{karras2019style}, videos~\cite{dvdgan,deepfakesurvey}, and audio~\cite{wang2017tacotron} that in some conditions are indistinguishable from those created by a human. Recent progress in generative models, such as those based on Generative Adversarial Networks (GANs) for images \cite{goodfellow2014generative}, and Transformer Architectures \cite{vaswani2017attention} for text, have to led to fears of these models being exploited for malicious use via Deep Fakes \cite{deepfakesurvey}, or the creation of machine-generated fake news at scale~\cite{fake-news}. 
In addition, the recent progress made in the development of neural architectures has resulted in models that can learn rich representations of data. They can thus be applied to a broad range of discriminative and generative tasks. For example, the same (or similar) model architectures trained on separate image datasets can be used to achieve good performance on biometric identification (e.g., face recognition), detection of cancerous regions in a mammogram \cite{shen2019deep}, or for classifying buildings and people from an unmanned vehicle \cite{human-object}. 

The context in which a model is applied can be far removed from that which the developers had intended, a major point of concern from the perspective of human-centered machine learning \cite{hcml}.  
However, there are few mechanisms that allow someone to share their work broadly while also restricting use in applications that may be of concern, such as large-scale surveillance or the creation of ``fake'' media. In some cases, the developers or technology creators may legitimately want to control the use of their work due to concerns arising out of the data that it was trained on, the technology's underlying assumptions about deploy-time characteristics, or the lack of sufficient adversarial testing and testing for bias. This is especially true of AI models that are difficult or expensive to recreate.  For example, given that models such as GPT-3  ~\cite{brown2020language} reportedly cost over \$10 million (U.S.) to train, very few organizations are positioned to train (and potentially, need to retrain) a model of similar size. 

The computing community has begun to respond to the conflict between powerful new technologies and the value of openness. A number of organizations have expressed concerns about incorrect or irresponsible use of AI and have proposed AI ethical guidelines (see an example from Microsoft \cite{microsoftrai2021}) and Responsible AI initiatives \cite{rao2019practical}. While such guidelines are useful and help shape policy, they are not directly enforceable, leading to fears of ``ethics washing'' \cite{ethics-washing1}. Governments have also taken note of the risks associated with certain types of AI applications and have passed legislation such as data protection laws in the European Union \cite{gdprtext2019} and San Francisco's Acquisition of Surveillance Technology Ordinance, banning the use of facial recognition technology \cite{Conger2019May}. While these are legally enforceable, government action requires prolonged deliberation by policymakers who often lack direct experience in computer science, let alone machine learning or artificial intelligence. Because development of regulations in the technology space tends to lag behind the development of new technologies, it is important for the computing research community to explore avenues for change that complement slower, top-down regulation. \footnote{As one example of such lag, the FTC sued Facebook in December 2020 for monopolistic conduct, including its 2012 acquisition of Instagram and 2014 acquisition of WhatsApp.}

In this paper we advocate for the use of {\em licensing\footnote{Our use of the word ``license'' is in the sense of a ``legal contract'' and not a ``permit'' or ``certificate''.}} as a mechanism for enabling {\em legally enforceable} responsible use. Licenses can help democratize the definition of responsible use by enabling AI developers and other stakeholders in AI systems (e.g., people who create data) to incorporate permissive- and restrictive-use clauses, based on {\em their} view of how {\em their} system(s) should or should not be re-purposed.
We further demonstrate how existing AI ethical guidelines may be operationalized through licensing. We argue that licenses could serve as a useful strategy for creating, expanding, and enforcing responsible behavioral norms as a complement to governmental legislation, and further serve to demonstrate that the computing community is itself committed to the responsible use of the technologies it creates. This approach is highly ``human-centered'' \cite{lee2020humancentered}, as it focuses on the interests of both developers and the general public.

Our paper makes the following contributions:
\begin{itemize}
    \item We discuss licensing as a mechanism for supporting the responsible use of AI, using behavioral use clauses.
    \item We present exemplar clauses that buttress concepts commonly found in widely-adopted AI ethical guidelines.
    \item We demonstrate how information from AI FactSheets~\cite{FactSheets} can be incorporated into a license for an AI system.
\end{itemize}

Together, these contributions advance a new approach to thorny responsible AI challenges.

\section{Related Work}
Here, we present a brief overview of the different approaches commonly used to promote the responsible use of AI. 

In reviewing this work, we emphasize the unique challenges AI systems create. Compared to traditional software systems, AI systems do not offer the same degree of control and run-time guarantees of system behaviour. While traditional software systems are also deployed in social systems that are highly complex and unpredictable, they can be systematically debugged, thoroughly tested, and versioned. In contrast, the output characteristics of AI systems are dependent on the {nature} of the dataset(s) used to train them, as well as the design and training procedure of the model. This makes it hard to be completely certain about how an AI model is likely to respond to known unknowns and unknown unknowns. While there is ongoing work that attempts to solve these problems for specific classes of models (e.g., ~\cite{ribeiro2018semantically} and~\cite{strobelt2018s}), there remain challenges, and there are no generalizable solutions.

\subsection{Best Practices for AI Development and Deployment} \label{sec:toolkits}
To help build more robust AI systems, researchers have built several  toolkits to study and address bias in datasets \cite{vasudevan20lift, lift,  bellamy2019ai}. In addition, behavioral testing frameworks~\cite{behavioral-testing} have been developed, including those that adversarially attack machine learning models to assess their sensitivity to input features \cite{art2018}, alongside various other tools\footnote{see e.g. Google's ``Playground'' and ``AI Explanations'': https://github.com/tensorflow/playground https://cloud.google.com/ai-platform/prediction/docs/ai-explanations} that help visualize the processing of inputs in an AI system. %

Technology companies and policy institutions have begun to release principles and guidelines for the use of AI.
According to a recent survey report \cite{ethics_survey} that studied $84$ AI Responsible AI and ethical guidelines, 73-86\% of the guidelines surveyed included principles of {\em Transparency}, {\em Trust and Accountability} and {\em Justice \& Fairness}. Principles of `Transparency' and `Trust' in guidelines include aspects related to the acquisition, use, and transformation of data~\cite{EU1,becks2021}, the model design \cite{ieee2021} and its end-use application \cite{cutler2019everyday}. Furthermore, some guidelines suggest that AI systems should be `Accountable and Fair' where they suggest that decisions of an AI system should be interpretable and explainable to a trained human worker \cite{montreal2021}. Additionally, some guidelines also recommend that AI systems should explicitly identify themselves as such \cite{montreal2021} and that decisions taken by an AI system, or with the assistance of an AI system, should have a human being accountable \cite{ibm2021}. 

How does one ensure that such principles are followed? Today, enforcement is based on self-regulation by technology developers to voluntarily comply with such guidelines. In other words, there is no external enforcement. This has led to fears that, in the absence of regulatory mechanisms that enforce the spirit of such guidelines, the reliance on self-regulation is resulting in shallow appearances of ethical behavior (colloquially  referred to as ``ethics washing'') \cite{ethics-washing1}.   






\subsection{Interpretability and Explainability for Responsible AI}
To understand the roles licenses might play in governing the use of AI, it is important to consider the varying degree to which models can provide interpretability regarding their decisions, and the role that interpretability and explainability research initiatives have played so far in the pursuit of `responsible' AI.

Depending on the underlying architecture used by machine learning models, the {\em interpretability} of downstream systems will vary. For instance, models based on Decision Trees allow users to study the decision made by each intermediate node in the tree; this can help determine the {\em reasoning} employed. Other models, such as those based on Lasso Linear Regression, can provide some explainability especially when used with easy to interpret features. Such models have been used in disease risk prediction and studies of clinical factors. \cite{bursac2008purposeful,turi2017peer}. Bayesian formulations are another class of machine learning models that can provide explanations using the conditional probabilities and dependencies between sets of variables \cite{koller2009probabilistic}. For example, such models are used to help indicate the likelihood of a patient having a particular disease given their medical symptoms and health parameters \cite{alexiou2017bayesian}.

In order to provide a degree of explainability in deep learning based AI systems, methods such as LIME \cite{LIME} offer locally interpretable, model-agnostic explanations by learning a feature based linear regression model trained to mimic (predict) the outputs of the deep learning model. The models are trained by generating samples after perturbing the instance under observation and the corresponding outputs. Other methods include SHAP \cite{SHAP} which also builds locally-explainable models, and uses techniques from cooperative game theory to model how different features contribute towards the final output. 
Instead of building simpler ML methods to model the output of black-box neural networks, developers have done this by enabling {\em visualizations} of model components - such as heatmaps of words \cite{lee2017interactive} or images \cite{xu2015show}. Probing (studying the effect of small changes in the input) \cite{niven2019probing} as well as, using counterfactuals \cite{byrne2019counterfactuals}, are other ways in which the neural network characteristics are studied.

\subsection{Documentation Efforts}
Better documentation for AI systems is another tool researchers investigated for advancing responsible AI. Providing details such as the data used for the development and testing of AI systems, the nature of testing and error characteristics, and risks of bias could help end-users determine the applicability or ``fit'' of a model for their needs. AI Factsheets~\cite{FactSheets} have introduced questionnaires seeking details about fairness and robustness checks and performance benchmarks, and they require developers to list intended uses in order to populate a summarized template sheet. Similar suggestions are included in AI Model cards \cite{ModelCards}, which advocate for quantitative evaluation to be broken down by individual, demographic, and other domain-relevant conditions. 

The datasets used for training and testing AI systems are key aspects to consider during development. The Datasheets \cite{Datasheets} for Datasets proposal introduces a questionnaire that captures the details about the purpose and nature of a system's data choices, including details around labeling, sources of noise, source of the dataset, who the creators are, and whether there might be relationships between data samples (e.g., social connections).

\subsection{Limited Release of AI systems}
The growing capability of AI systems, as well as the potential for their misuse, have made creators of AI systems increasingly concerned about sharing their work.  OpenAI, for example, only selectively disclosed elements of GPT-2 when it released GPT-2 \cite{radford2019language}: ``{\em Due to concerns about large language models being used to generate deceptive, biased, or abusive language at scale, we are only releasing a much smaller version of GPT-2 along with sampling code. We are not releasing the dataset, training code, or GPT-2 model weights.''}\footnote{https://openai.com/blog/better-language-models/}  OpenAI was similarly circumspect in releasing GPT-3 by providing researchers with access to GPT-3 only through its own cloud-hosted API. Other AI technology providers such as IBM, Amazon and Microsoft recently announced \cite{Magid2020Jun} they will no longer offer general-purpose facial recognition technology to law enforcement agencies. 
	
 Not all AI technology providers want to ``hold back'' their technology from potential users, and not all AI technology providers are positioned to release only bits and pieces. 
 For those technology providers who wish to release their work, and yet simultaneously impose behavioral constraints on future users, licensing appears to provide a pathway that is less extreme than limited-releases and price-based mechanisms that create a barrier to entry \cite{synced2020}. 

 In the next section we provide an introduction to licensing as a legal framework. We then describe exemplar clauses based on some ethical AI principles and information shared in AI Factsheets~\cite{FactSheets}. These proof-of-concept exemplars help to demonstrate how the use of behavioural-use clauses could offer an enforceable mechanism for promoting the responsible use of AI,  grounded in the existing legal framework for protecting intellectual property (``IP'').

\section{Licensing as a Legal Framework}

Let us continue by establishing the basis for the use of licensing as a mechanism for controlling end-user behavior.  We describe the structure of a license and the key elements for making it effective.
Governments have created intellectual property rights
in order to give authors and inventors control over the {\em use} and {\em distribution} of their ideas and inventions. Intellectual property (IP) rights thus act as an incentive for creators to engage in creative activities (with a potential cost of limiting the diffusion of ideas). Examples of such intellectual property rights include patents, copyrights, and trademarks. By offering new licensing opportunities to AI developers and stakeholders (e.g. data creators), these groups can take advantage of existing IP frameworks.

\subsection{Background}
In general, a {\em license} is a contractual agreement that grants {\em permission} by a qualified authority, such as the owner of IP rights, to a licensee. 
Licenses may have territorial limits, temporal limits, or other restrictions.  These permissions are, of course, limited by the licensor's power to confer such permissions. When IP owners license the use of their IP to others, they forfeit their right to exclude those licensees from using the IP. 
IP owners benefit from licensing because they maintain ownership over their IP while profiting from the licensee's use of their IP, either in the form of royalties or in the co-development of new IP. 

IP licenses are formed contractually in the form of license agreements. The terms of a license are typically dictated by those who own or control the IP. ``IP'' can include many different intangible rights -- e.g. patents (inventions), copyright (works of authorship including technical manuals, software, specifications, formulae, schematics, and documentation), know-how (e.g. expertise, skilled craftsmanship, training capability), trade secrets (a protected formula or method, undisclosed customer or technical information, algorithms, etc.), trademarks (logos, distinctive names for products and services), industrial designs (the unique way a product looks such as a computer’s molding), and semiconductor mask works (the physical design of semiconductor circuits).

Licenses may be for certain IP rights only (e.g. a license to practice an identified patent or to copy and distribute a certain work). Licenses may be for all the IP rights that are necessary to sell, market, and/or use a specific type of technology (e.g. a license to develop a new widget that is created from a patented process and has a distinctive design).

Thus, an IP owner could permit users to create derivative works but only with attribution to the original author (e.g., Creative Commons License).  Alternatively, an IP owner could require users to refrain from using the licensed technology in a particular geography or in a particular type of application.  

In this paper, we seek to encourage entities and individuals who create AI tools and applications, to leverage the existing IP license approach to restrict the downstream use of their tools and applications (i.e., their ``IP''). Specifically, IP licensors should allow others to use their IP only if such licensees agree to use the IP in ways that are appropriate for the IP being licensed.
While contractual arrangements are not the only means to encourage appropriate behaviour, it is a mechanism that exists today, is malleable to different circumstances and technologies, and acts as a strong signaling mechanism that the IP owner takes their ethical responsibilities seriously. Put another way, licenses are one readily available tool that developers and technology creators can use \textit{today} to take a step towards responsible AI.

While there are benefits of licensing regimes, one must also be mindful of the fact that violations of the licensing terms are enforced by the IP owner (or its designee) via legal action in the form of litigation. At least two causes of action could be brought in this context -- breach of contract and infringement of IP rights.  Thus, those who opt for a licensing regime to enforce responsible behavior in the AI space must be prepared to enforce such terms using available dispute resolution mechanisms.  In the US, those mechanisms are primarily the judicial system and private arbitration.

\subsection{Structure of a License} \label{sec:license-structure}
	The basic structure of a license requires, at minimum, the following:
	
	\begin{itemize}
	    \item Licensor: A person, business or organization with exclusive legal rights over the assets,
	    \item Licensee: A person, business or organization that has been granted legal permission by the licensor, 
	    \item Identification of the subject Intellectual Property, and
	    \item Restrictions:  geographic, temporal, behavioral, ability to sublicense, royalty terms, etc.
	\end{itemize}
    
    Multiple examples of behavioral restrictions exist in common software licenses, such as limiting the licensee's access for `internal use' only; prohibiting modification of the software code; and restricting use of the software to specific countries. These license restrictions can be tailored for each type of technology being introduced. For example, in the novel field of ``robot law'', at least one set of ``ethical license terms'' has been proposed with restrictions requiring default privacy settings and a kill switch \cite{cooper2016application}. 
    Such behavioral restrictions generally originate with the licensor, and may or may not be subject to negotiation between the licensor and licensee. 

	An IP owner who wishes to act as an IP licensor should be cognizant of the following three questions in monitoring compliance with the license terms: 
    (1) How can I detect license term violations? 
    (2) What are the potential consequences of such violations?
    (3) What are the available enforcement mechanisms for the license terms?
    
\subsubsection{Detecting Violations}
How does an IP licensor detect when a licensee has exceeded the permissible scope of the license?  That is, will it be obvious to the IP licensor if or when a licensee of software that is licensed for ``internal use only'' is shared by the licensee with third parties? Certain types of IP are more readily policed than others. For example, it would arguably be much more difficult for a licensor to detect third parties' unlicensed use of spelling-autocorrection software as opposed to third parties' unlicensed use of an online photo.  Depending on the difficulty of detecting violations of IP license terms, the licensor may wish to incorporate other clauses in the contract to ensure compliance, such as annual audits or periodic reports of IP usage. Standardization of license components reduces the burden on developers when it comes to releasing assets. A good example of this is the Creative Commons License which provides a set of options for the licensor to choose from about how the assets can be used by the licensee. 
    
    \subsubsection{Consequences of Violations}
   The potential consequences of breaching the license terms vary widely. In some instances, the breach -- especially if it was made inadvertently -- may be corrected by a simple notice from the licensor that the licensee has breached the contract. Other breaches can be catastrophic, such as exposing licensor's source code, which can dramatically limit the ability for a licensor to control the use of that code.  A licensor should think through the possible ranges of consequences for different types of breaches, and be prepared to identify ways for each type of breach to be cured.  This further determines how other clauses in the license agreement may be drafted. 
    
    \subsubsection{Enforcement}
    The available enforcement mechanisms can influence the licensees' willingness to stay strictly within the license terms.  Some IP licensors have technical mechanisms to enforce their license terms. In the event a licensee breaches the license terms, the licensor can cut off access to its IP.  This is feasible, for example, where the IP licensor owns a digital platform (such as a Mobile App Store) or provides software as a service via the cloud.  Other IP licensors, who lack such technical mechanisms of enforcement, must rely on the threat of legal action to effect compliance with the license terms. In the United States and in other countries with robust judicial systems, an IP licensor can rely on its lawyers and the court system to enforce the rules dictated by its license terms.
    
     We note that apart from enabling enforcement, license restrictions may also serve as a deterrent; corporations and organizations, including those tasked with the deployment of high risk AI systems, have stringent legal reviews of tools and software for the license terms associated with them. This includes open source software where some flavours of licenses are viewed more favourably than others with regard to commercialization. Thus, any restrictions on use included in licenses and contracts are likely to be self-enforced to avoid the risk of expensive litigation.

\subsection{Lessons from Open Source Software}

 The advent of open source software in the 1990s
 popularized the notion that software should be developed collaboratively and freely shared within the developer community.  The Open Source Definition \cite{ContributorstoWikimediaprojects2021Sep} presents a mature open-source philosophy and defines the terms of use, modification, and redistribution of open-source software. These terms include the following: (1) free redistribution of source code; (2) modifications are permitted and must be distributed under the same terms as the original license; (3) no discrimination against any person or groups of persons; (4) no discrimination against fields of endeavor; and (5) technology neutrality. Today, most if not all participants in the software development community are familiar with open source ideas, under which software licenses grant rights to other users which would otherwise be reserved to the writer of such software under copyright laws.
  One of the popular types of restrictions is the ``Do No Evil" restriction, popularized by JSON (JavaScript Object Notation), a format for storing and transporting data between servers and web pages:
 \textit{``Permission is hereby granted, free of charge, to any person obtaining a copy of this software and associated documentation files (the ``Software"), to deal in the Software without restriction, including without limitation the rights to use, copy, modify, merge, publish, distribute, sublicense, and/or sell copies of the Software, and to permit persons to whom the Software is furnished to do so, subject to the following conditions:
  The above copyright notice and this permission notice shall be included in all copies or substantial portions of the Software.
  \textbf{The Software shall be used for Good, not Evil}.''}
 
The last sentence is understood to violate the open source definition, given that it restricts usage. 
Furthermore, most users are likely able to convince themselves that they are not engaged in evil activities, and that they are thus compliant even with the ``Do No Evil" terms of a JSON license. This example highlights challenges in creating enforceable licenses.

\section{Case Studies}

In contrast to the the highly developed community norms around open source, discussions of AI negative impacts lack standards. While academics have pointed out the need for new frameworks in this growing space (see, e.g., the case for ``data justice'' \cite{datajustice}), the current vacuum has yet to be filled with government regulation. To work towards licensing as a topic of norm-building, we present two case-studies where we adapt some recommendations from AI ethical guidelines as well as AI FactSheets \cite{FactSheets}, as license clauses. The goal here is to illustrate the types of clauses that \textit{could} exist and not necessarily argue for or against these specific clauses. In other words, a key contribution of these case studies is to make the point that licenses for responsible AI are highly plausible and achievable in the near-term.

 \subsection{Case Study - Ethical Guidelines} \label{sec:case-study-guidelines}

	As previously stated, a number of companies and countries have released ethical guidelines for the use of AI, with many overlapping concepts.
	How do companies, researchers, and developers of AI systems -- all in the position of a technology licensor -- implement these guidelines?  In this section, we list a few elements from the {\em Montreal Declaration on Responsible AI} \cite{montreal2021} and show how each guideline listed below can be operationalized in license clauses. 

   
    \subsubsection{Propaganda and False Information}
    ``Artificial Intelligence systems (AIS) must not be developed to spread untrustworthy information, lies, or propaganda, and should be designed with a view to containing their dissemination'' (Principle 2, Statement 5).
    
     \noindent {\bf Example Clause:} ``{\em Licensee will not use the Licensed Technology to enable the distribution of untrustworthy information, lies or propaganda, and if Licensee discovers that such distribution is unintentionally occurring, Licensee will put in place countermeasures, including human agents, to prevent or limit such distribution.}''

    \subsubsection{Imitation of Human Characteristics}
    ``The development of AIS must avoid creating dependencies through attention-capturing techniques or the imitation of human characteristics (appearance, voice, etc.) in ways that could cause confusion between AIS and humans'' (Principle 2, Statement 6).

    \noindent {\bf Example Clause:} ``{\em The licensee will not use the Licensed Technology in a manner that would imitate human characteristics and cause third party confusion between AIS and humans.}''

    \subsubsection{Damage to Reputation and Manipulation}
    ``The integrity of one's personal identity must be guaranteed. AIS must not be used to imitate or alter a person's appearance, voice, or other individual characteristics in order to damage one's reputation or manipulate other people.'' (Principle 3, Statement 8)

    \noindent {\bf Example Clause:} ``{ \em The licensee will not utilize the Licensed Technology in applications that imitate or alter a person's likeness, voice, or other identifiable characteristics in order to damage his/her reputation.}''

    \subsubsection{Transparency} 
    ``Any person using a service should know if a decision concerning them or affecting them was made by an AIS.'' (Principle 5, Statement 8)

    \noindent {\bf Example Clause:} ``{\em The licensee will disclose to an end-user if and when use of the Licensed Technology affected a decision such end-user, including by stating whether the Licensed Technology itself is an AI.}'
    
    As can be seen, in the examples above, statements in AI ethical principles or guidelines can be transformed into license clauses and the use of such clauses can help {\em enforce} the principles laid out in such guidelines. How such license terms are phrased can vary widely, based upon the goals of the licensor, its industry, the type of technology being licensed, the potential for misuse or harmful applications, and the ease of enforceability. We include an exemplar license based on some of these usage restrictions in Section \ref{sec:exemplar-license}. 

 \subsection{Case Study - FactSheets}
 AI FactSheets~\cite{FactSheets} are a form of Supplier's Declaration of Conformity (SDoC) that can be released by AI developers to describe the characteristics of the associated product or service. FactSheets may include sections that discuss the {\em robustness} of models to adversarial attacks, {\em bias} due to datasets used, {\em Optimal} and {\em Poor} operating conditions, {training and test data use for evaluation}, its domain of application etc. 
 We consider the five examples available on the IBM AI FactSheets webpage\footnote{https://aifs360.mybluemix.net/examples/} to demonstrate how clauses may be created. The models are: (i) Audio Classifier, (ii) Object Detector, (iii) Image Caption Generator, (iv) Text Sentiment Classifier, and (v) Weather Forecaster. 
 
 To present exemplar clauses derived using AI FactSheets we use one or more of the following fields: (i) Intended Domain (ii) Performance Metrics (iii) Bias (iv) Robustness (v) Domain-Shift (vi) Optimal Conditions (vii) Poor Conditions
 \subsubsection{Audio Classifier}
 The Audio Classifier\footnote{https://aifs360.mybluemix.net/examples/max\_audio\_classifier} classifies an audio stream into day-to-day sounds such as ``outdoors'', ``speech'', ``music'' etc. It has been trained using the AudioSet Dataset \cite{audioset} and uses a multi-level attention based classifier \cite{audiosetmodel}. The AudioSet dataset used to train the model contains a vast majority of audio files from YouTube that contain speech and music. The factsheet reports that a study of different voice types and music genres has not been performed. The factsheet further reports that the classifier only performs well when the audio file contains only one or two distinct classes and the file does not have a high degree of noise.
 
 Given the information in the factsheet and the limitations of the model, one could specify a clause that requires adhering to optimal operating conditions, as well as disallowing the use of the classifier for legal and security applications, due to the sensitivity to noise and the risk of bias in voice-types. 
 
 \noindent {\bf Example Clause:} ``{\em Licensee will not utilize the Licensed Technology in applications that involve the identification or classification of human beings for legal or security purposes if the intent of such use is to impact such human beings' livelihoods or personal liberty.}''
 
 \subsubsection{Object Detector}
 The Object Detector\footnote{https://aifs360.mybluemix.net/examples/max\_object\_detector} recognizes $80$ different classes of objects from the COCO dataset \cite{coco} using the SSD MobileNet v1 Model \cite{mobilenet}. The factsheet reports that the trained model has been studied for gender bias but has not been evaluated for other forms of bias. The model is sensitive to the image patterns seen during training and also exhibits performance degradation when image transformations add noise. Further, the model may not detect objects or may classify them if that object-type was not seen during training (as opposed to identifying an object with generic class-label `thing'). Finally, poor resolution and lighting also results in low performance.
 
Given the information provided in the factsheet, a license for a service using this model could restrict its use in autonomous agents due to the nature of errors and sensitivity to lighting, and also prohibit use on security camera footage (which often has poor quality and could also contain low-light footage). In addition, a clause could also be included that introduces certain {\em compliance} requirements - such as requiring derivative applications to prominently display the limited set of object classes a system may be used for. 
 
 \noindent {\bf Example Clause:} ``{\em Licensee will not use the Licensed Technology in the context of law enforcement, security, or surveillance.   In the event that Licensee sublicenses any of the Licensed Technology in the permitted uses contained in the original license, Licensee will prominently display in its license terms all of the limitation of its system, including those contained in this license.}''
 
 \subsubsection{Image Caption Generator}
 This model\footnote{https://aifs360.mybluemix.net/examples/max\_image\_caption\_generator} generates a caption using a limited vocabulary to describe the contents of an image. It has been trained using the COCO dataset and uses the Inception v3 model \cite{inception}. As in the previous model, this model is known to have exhibited gender biases due to the dataset. The model does not perform well on low-light footage and always makes a prediction even if the object was not something seen in training data.
 
 Thus, given the information in this FactSheet, a clause could prohibit use in assistive technologies for the visually-impaired which could incorrectly describe a scene in a way that could endanger lives.

 \noindent {\bf Example Clause:} ``{\em Licensee will not use the Licensed Technology in order to perform a visual identification function for humans that are visually impaired.''}
 
 \subsubsection{Text Sentiment Classifier}
 The Text Sentiment Classifier\footnote{https://aifs360.mybluemix.net/examples/max\_text\_sentiment\_classifier} is based on the BERT pre-trained model and it has been fine-tuned on a corpus of annotated sentences from Wikipedia. It classifies a sentence based on its sentiment polarity (positive/negative). The Factsheet indicates that the model may express racial and gender bias. This model works well when sentences are well structured like formal Wikipedia articles and express sentiment unambiguously. The model does not work well on sentences that includes elements like sarcasm, passive-aggressive statements, multiple polarities, or socially or racially offensive language.
 
 Given the operational characteristics of the model, clauses that restrict its use in automated content moderation would be applicable. It may also make sense to restrict the use of models for applications that interact with text data that are quite different from Wikipedia articles (e.g. conversational applications, social media).
 
 \noindent {\bf Example Clause:} ``{\em Licensee will not use the Licensed Technology in connection with the identification of gender or race.  Further, Licensee agrees to supplement the use of Licensed Technology with human intermediaries in contexts involving informal speech (such as sarcasm or idioms) or unfiltered social media (including but not limited to Twitter).''}

 \subsubsection{Weather Forecaster}
 The Weather Forecaster model\footnote{https://aifs360.mybluemix.net/examples/max\_weather\_forecaster} uses weather-related time series data to predict temperature, visibility, dew point, humidity, wind speed, etc. The models have been trained using data from the northeastern US. The factsheet reports that the model has very poor prediction performance on non-northeastern US weather data.
 
 Given the limitations of this model, clauses could be defined to prevent use on data from other regions especially in applications that could put lives at risk, e.g. alerts for air/sea navigation or storm alerts for evacuation.
 
 \noindent {\bf Example Clause:} ``{\em Licensee will not use the Licensed Technology to provide weather prediction in any geographic location outside of the northeastern sector of the United States.''}
 
 As can be seen, information described in AI FactSheets can be helpful in defining system-specific license clauses. Recently there have also been calls for AI researchers to consider the negative impacts of their work \cite{hecht2018s}. Some AI conferences such as NeurIPS now require authors to consider the potential broader impacts of research projects.  We note that, similar to AI Factsheets and Ethical Principles, such impact statements could also be used to inform restrictions or appropriate-use clauses in licenses. 
 

\subsection{Licensing Data}

The performance of AI systems is dependent on the nature of training dataset(s) and the accessibility of such datasets (i.e., ``Openness"). In our discussion so far, we have considered the primary IP being licensed to be AI models. However, datasets are routinely licensed as stand-alone IP\footnote{See e.g. data from Facebook and Twitter: https://developers.facebook.com/docs/graph-api/overview/, https://developer.twitter.com/en/docs/tutorials/consuming-streaming-data}.
Many widely-used datasets contain errors, reflect and reinforce problematic societal biases, and spread misinformation \cite{paullada2020data} -- any of which can create opportunities for harm.

Due to the growing potential for irresponsible use of datasets by AI systems, there is an accelerating effort within academic communities to develop mechanisms that restrict how datasets are collected, used, and shared. One example mechanism is a Data License Agreement (``DLA''), which is a specific type of license that defines the arrangement of data exchange between a licensor and a licensee.  In a typical scenario, the licensor will seek to limit use of a dataset by the licensee by including provisions in the DLA that identify: 1) a specifically tailored definition of the dataset being licensed; 2) specific persons permitted to use the licensed dataset (i.e., the licensee); and 3) specific purpose(s) for which the licensed dataset may be used. For example, the Montreal Data License Generator \cite{MDL} uses a questionnaire to generate intellectual property (IP) licensing terms that can be attached to datasets to govern its distribution. The framework provided by the Montreal Data License authors provides a well developed starting point for attempts to use data licensing to drive responsible AI efforts. As another example, datasets hosted by Stanford's 
Center for Artificial Intelligence in Medicine and Imaging requires end users to agree to a `Research Use Agreement' (e.g. \cite{irvin2019chexpert}).

Similar to other IP agreements, DLAs could be extended with responsible use clauses, or behavioral use restrictions.
Doing so could further minimize the potential for irresponsible or harmful use, while simultaneously maintaining the ability to derive value from the dataset. The introduction of responsible use clauses or behavioral use restrictions into DLAs may also encourage compliance with various state and federal regulations (e.g., data privacy regulations) governing the collection, use, and disclosure of certain data, such as personal information.  Implementation of DLAs is also in line with efforts to support \textit{data justice}, i.e. efforts to achieve ``fairness in the way people are made visible, represented and treated as a result of their production of digital data'' \cite{datajustice}. Introducing responsible use clauses helps to ensure that the downstream uses of datasets reflect the values of those who produce data. 

There are several challenges that will need to be addressed to support early adoption of data license practices. First, some of the most important datasets are already created under various platform-dependent legal contexts. For instance, Wikipedia, which is widely used in AI research \cite{brown2020language,vincent2021deeper}, is licensed as CC-BY-SA. This means that Wikipedia contributors are not able to decide upon a new license to control how the data they produce is used; the choice has already been made by the Wikipedia community. Similarly, the digital records produced when people use popular technologies like search engines and social media would be difficult to license. Currently, many of the people who help produce such data (e.g. by using a search engine) may not even realize they are generating valuable data, and companies would have no incentive to offer the public the ability to license such data. 

Another challenge is the possibility that data licenses will conflict with licenses chosen by AI developers. That is, a ``downstream'' AI model may have a more permissive license than an upstream dataset.
Negotiating these kinds of conflicts will be an important challenge for proving the viability of the data license concept.

An ideal solution to license conflicts would be for AI developers and stakeholders to support democratic governance approaches to deciding on licenses and clauses. For instance, a system that is highly dependent on data from a particular community (e.g. Wikipedia, GitHub, Mechanical Turk workers) could give interested data contributors the ability to vote on various aspects of the behavioral use license or otherwise participate in a deliberation process. This approach would be highly in line with calls to empower people who do ``data labor'' to generate valuable data \cite{arrieta2018should,vincent2021deeper,vincent2021data}.

\section{Limitations and Open Questions}
We propose licensing as one additional tool in the growing toolkit of sociotechnical mechanisms available for actualizing responsible AI goals. Licensing has attractive properties, such as providing researchers and developers with more control over how their artifacts are used while potentially maintaining many of the desirable properties of openness. However, there are several challenges to, and limitations of, licenses as an approach that must be addressed before licensing can be a maximally effective. 
 
 \noindent \textbf{Adoption.} Perhaps the most significant challenge facing licensing as a tool for responsible AI is adoption. There are many different types of licenses that exist and norms around how they are used. For instance, in the case of open sourced systems, studies have shown that the adoption of more {\em permissive} licenses leads to an increase in community contributions and might even result in abandonment of competing projects that use more restrictive licenses \cite{ICSE10}. A survey on the use of open source licenses by \citeauthor{toughuselicense} found that developers currently have strong intrinsic beliefs that affects their choice of licenses. Developers also face difficulties in understanding licensing terms as well as in dealing with incompatible licenses \cite{toughuselicense}.  
 
 The value of behavioral use clauses must therefore be communicated to those releasing software. The clauses must also not conflict with the other value people see in licenses (e.g., permitting or restricting redistribution of code, etc.). To help address some of these challenges, it may be helpful to consider the development of licenses that correspond to existing ethical guidelines. Creating a repository of such licenses could help potential licensors select existing pre-defined licenses that align with the principles to which they want to adhere. Further, the creation of such community assets could help with adoption by easing the burden of drafting legally sound license clauses. 
 
 In addition to the creation of such repositories, licenses could also be made modular using license generators wherein AI developers could select clauses (or ethical principles) as well as other license elements, such as the terms of commercial/non-commercial distribution, description of penalties or conditions of violation etc, that they would like to apply with the release of their AI systems. License generators would also help the groups who develop AI systems to navigate challenging consensus-finding discussions about which specific uses should be restricted. Furthermore, license generators could help the community coalesce around relatively simple licenses, to mitigate potential effects where complicated licenses are difficult to enforce. As noted above, an ideal system would incorporate the interests and perspectives of all stakeholders, including people who contribute data.
 
 \noindent \textbf{Incompatibility with Open Source.}
 Certain open source licenses such as GPL V3\footnote{https://www.gnu.org/licenses/gpl-3.0.en.html} require that freedoms received much be passed on when software or code is modified and/or re-licensed. This implies that any additional restrictions on use or distribution that were not originally present in the GPL license cannot be added. Licenses with responsible-use clauses or usage restrictions can therefore not be applied to any software or code that was originally distributed with licenses such as GPL V3 making them incompatible. On the other hand, other open-source licenses such as Apache 2.0 are more liberal and permit re-licensing of derivative works under new clauses. IP lawyers in software companies routinely scrutinize licenses when permitting use of software or source code for development (open-source or otherwise), and licenses with usage restrictions would now need to be subject to the same legal scrutiny.

 \noindent \textbf{Enforcement.} Licenses are only
 useful if they are enforced. As discussed earlier, licenses include consequences of breaching the terms of contract. Similar to ethics washing via ethical guidelines and principles~\cite{ethics-washing1}, if licenses are not enforced they run the risk of accentuating that problem. In some instances terms may be difficult to enforce -- for instance if a large corporation violates the responsible use clauses of an AI system shared publicly by an independent developer (with limited resources) using a GitHub Repository. Furthermore, more complicated licenses will be more difficult to enforce. We note that these problems are not unique to our approach and apply to licensing in general including open-source licenses. 
 
 For AI systems offered as cloud based APIs (or similar) termination of service could serve as an easier mechanism of enforcement as opposed to litigation which can be time consuming and expensive. This means an API-based approach could provide an appealing approach to enforcement.

To support AI developers with more limited means, researchers might try to bootstrap collaborative communities that publicly list violations along with details such as the date, the nature of violation, and response received. Beyond `naming' and `shaming' which may have limited efficacy, such communities could engage the services of pro-bono legal initiatives to help enforce terms of breaches collectively (if one entity/organization/individual has been named in multiple violations or if one clause of a license instance has been violated by multiple parties), based on an assessment of the merits of the case, the severity of impact of violation etc. 
 
Scaffolding support for AI developers will be critical, because enforcement will inevitably create new responsibilities and labor for developers, many of whom are volunteers and likely to have limited resources for enforcement. If leaders of the responsible AI initiatives can lower this burden, a licensing approach is likely to be more effective. 
 
 

 
 
\noindent \textbf{Customization.} There are many possible behavioral use limitations people may desire and therefore heavily customized licenses may seem attractive.  However, this could lead to a complex landscape of different behavioral licenses that might be confusing to licensees and require extensive conversations with legal professionals. Apart from creating clause/license repositories, as suggested previously, it may be useful to consider the development of community {\em standards} that define and standardize license elements to be used in AI license. Drawing parallels from the Open Source community - while there are hundreds of licenses hosted on the Software Package Data Exchange webpage \cite{spdx}, there are a few open source licenses that have managed to become very popular. We hope that similar community initiatives along with the development of AI licensing standards could help address some of the challenges discussed in this section.  
 

\noindent \textbf{Reinforcing Existing Power Structures.} 
Many legal frameworks reinforce or support existing power structures.  Do license agreements provide a way for people to say ``no''?  We would argue that licenses in the form we propose do not provide a direct mechanism for people to say ``no'', while still receiving the benefits that the code or data being distributed may offer. Therefore, people that decline to agree to these terms may be disadvantaged as a result. We recognize that this inability to say ``no'' can lead to people being deprived of their will~\cite{ahmed2017no} and that the politics of refusal~\cite{garcia2020no,manifest-no} are very important to consider. Licenses do not directly address these issues, and may be viewed as entrenching them. However, we would also argue that behavioral use licenses for code and data do acknowledge that these assets are in need of interpretation (the importance of which is highlighted in~\cite{manifest-no}) and that restricting certain behaviors can be viewed as a way of helping make that interpretation more transparent and concrete. 

\section{Conclusion}

Distribution of code, data, models, and access to APIs leaves researchers and developers in a conundrum. On one hand, this distribution fosters democratization, verification, reproduction, and advancement of scientific knowledge. On the other hand, these assets could be used for purposes that the developers never intended; some of these might be harmful or irresponsible. 
These issues are compounded in light of large-scale pre-trained models (such as GPT-3) that can be used for many downstream applications but require vast amounts of training data and computational resources. Sharing access is the only way that people would be able to build on such a system in practice.
Given that there is a well-established framework for IP licensing that exists today, we believe that licensing could be a powerful tool for developers in preventing negative impacts of technologies immediately. The example clauses we have identified apply to a range of different types of AI or machine learning systems. 
We believe that normalizing the use of behavioral restrictions via licenses will encourage responsible use of powerful AI tools and systems by downstream users, while recognizing that there are limitations in enforcing such terms. 

\section{Exemplar License} \label{sec:exemplar-license}
This section presents a license that has been adapted from the Apache 2.0\footnote{\url{https://www.apache.org/licenses/LICENSE-2.0}} license to include usage restrictions. The 
``Definitions'' standardize language throughout the license, while the clauses in ``Restrictions''
include some examples of restrictions from our case study in Section \ref{sec:case-study-guidelines}. 
As this example shows, licensing frameworks are fairly flexible and they can support the inclusion of custom use-case and software-specific clauses with ease. The section on restrictions is prefaced with text which ensures that downstream applications as well as other modification/use/redistribution continues to include the restrictions. This approach is similar to the one taken by open source licenses such as GPL V2\footnote{\url{https://www.gnu.org/licenses/old-licenses/gpl-2.0.en.html}} to ensure redistribution/re-use remains open-source. ``Terminations'' 
show requirements to be fulfilled in the event of a breach of the license terms. The remaining sections includes text that is typically present in software licenses to minimize risk arising from legal liabilities. 

\noindent Exemplar license text below:
\vspace{0.2cm}

{\footnotesize

\noindent This license governs the use of the accompanying software. If you access or use the software, you accept the License.  If you do not accept the License, do not access or use the software.    

\subsection{Definitions} \label{sec:license-def}

As used in this License, the following capitalized terms have the following meanings: 
\begin{enumerate}[(i)]
    \item ``License'' means the terms and conditions for use, reproduction, and distribution as defined by Sections one (1) through eight (8) of this document.  

\item ``Licensor'' means the copyright owner or legal entity authorized by the copyright owner that is granting the License.

\item  ``You'' (or ``Your'') means an individual or legal entity exercising permissions granted by this License.

\item The terms ``reproduce'', ``reproduction'', ``derivative works'', and ``distribution'' have the same meaning here as under U.S. Copyright Law. 

\item “Contribution” means the original software, additions to the original software, modifications to the original software, or derivative works of the original software.  

\item ``Contributor'' means any person or Licensor who provides a Contribution.  

\end{enumerate}

\subsection{Grant of Rights}

Subject to this License, each Contributor grants You a non-exclusive, worldwide, royalty-free copyright license to reproduce its Contribution, prepare derivative works of its Contribution, and distribute its Contribution or any derivative works of its Contribution that You create.

\subsection{Restrictions} \label{sec:license-restrictions}
\begin{enumerate}
    
\item        If You distribute any portion of the Contribution, You must include a complete copy of this License with the distribution; and

\item         You agree that the Contribution, or any derivative work of the Contribution, will not be used by You or any third party subject to Your control, to:

\begin{enumerate}[(i)]
\item enable the distribution of untrustworthy information,
lies or propaganda, and if You discover that such distribution is unintentionally occurring, You will put in place
countermeasures, including human agents, to prevent or limit such distribution.

\item imitate human characteristics
and cause third party confusion between Artificial Intelligence Systems and humans

\item imitate or alter a person’s
likeness, voice, or other identifiable characteristics in order to damage his/her reputation

\item {\dots} [More restrictions could be added.]

\end{enumerate}

 \end{enumerate}

\subsection{Termination} \label{sec:license-termination}

Upon the occurrence of any of the restricted uses listed above in ``3. Restrictions'', Licensor shall have the right to:
\begin{enumerate}[(i)]

\item terminate this License Agreement and disable any Contribution either by pre-installed or then installed disabling instructions, and to take immediate possession of the Contribution and all copies wherever located, without demand or notice;

\item require You to immediately return to Licensor all copies of the Contribution, or upon request by Licensor destroy the Contribution and all copies and certify in writing that they have been destroyed;   
\item  for a period of 1 year, provide a prominent notice on the Licensor’s website indicating that this License was violated by the Licensor;

\item  release/delete any and all data collected through use of the Contribution; and    

\item notify all parties affected by use of the Contribution.      
\end{enumerate}
Termination of this License Agreement shall be in addition to and not in lieu of any other remedies available to Licensor.  Licensor expressly reserves the right to pursue all legal and equitable remedies available under the law.

\subsection{Disclaimer of Warranty}

Unless required by applicable law or agreed to in writing, Licensor provides any Contribution (and each Contributor provides its Contributions) on an "As-Is" basis, without WARRANTIES OR CONDITIONS OF ANY KIND, either express or implied, including, without limitation, any warranties or conditions of TITLE, NON-INFRINGEMENT, MERCHANTABILITY, or FITNESS FOR A PARTICULAR PURPOSE. You are solely responsible for determining the appropriateness of using or redistributing a Contribution and assume any risks associated with Your exercise of permissions under this License.

\subsection{Limitation of Liability}

In no event and under no legal theory, whether in tort (including negligence), contract, or otherwise, unless required by applicable law (such as deliberate and grossly negligent acts) or agreed to in writing, shall any Contributor be liable to You for damages, including any direct, indirect, special, incidental, or consequential damages of any character arising as a result of this License or out of the use or inability to use any Contribution (including but not limited to damages for loss of goodwill, work stoppage, computer failure or malfunction, or any and all other commercial damages or losses), even if such Contributor has been advised of the possibility of such damages.

\subsection{Accepting Warranty or Additional Liability}

While redistributing the Contribution, You may choose to offer, and charge a fee for, acceptance of support, warranty, indemnity, or other liability obligations and/or rights consistent with this License. However, in accepting such obligations, You may act only on Your own behalf and on Your sole responsibility, not on behalf of any other Contributor, and only if You agree to indemnify, defend, and hold each Contributor harmless for any liability incurred by, or claims asserted against, such Contributor by reason of your accepting any such warranty or additional liability.

\noindent END OF TERMS AND CONDITIONS
}


\begin{acks}
We would like to thank Stephen Ibaraki for his continuous support during the course of this work. We would also like to thank Francesca Rossi, Michael Hind, Sriram Raghavan, and Jim Spohrer for their helpful feedback and suggestions. 

Funding/Support: The authors declare no additional sources of funding.
\end{acks}

\bibliographystyle{ACM-Reference-Format}
\bibliography{refs}

\end{document}